\documentclass[10pt]{IEEEtran}
\title{Sinusoidal analysis of memristor bridge circuit: rectifier for low frequencies}
\author{Oliver Pabst, Torsten Schmidt\IEEEauthorblockA{\\ Faculty of Electrical and Computer Engineering, \\ Technical University Dresden, \\ Dresden Germany, \\ oliver.pabst@mailbox.tu-dresden.de}}

\usepackage{mathptmx}
\usepackage[latin1]{inputenc}
\usepackage{graphicx}
\usepackage{amssymb}
\usepackage{amsmath}
\usepackage{pdfpages}
\usepackage{float}
\ifCLASSOPTIONcompsoc
 \usepackage[tight,normalize,sf,SF]{subfigure}
 \else
  \usepackage[tight,footnotesize]{subfigure}
  \fi
\newcommand*\Ron{\mathrm{R_{on}}}
\newcommand*\Roff{\mathrm{R_{off}}}
\newcommand*\IM{\mathrm{I_M}}
\newcommand*\VM{\mathrm{V_M}}
\newcommand*\IS{\mathrm{I_S}}
\newcommand*\VS{\mathrm{V_S}}
\newcommand*\fcut{\mathrm{f_{cut}}}
\newcommand*\tcut{\mathrm{t_{cut}}}
\newcommand*\Num{\mathrm{Num}}
\newcommand*\Den{\mathrm{Den}}

\newcommand*\fW{\mathrm{f_W}}
\newcommand*\RL{\mathrm{R_L}}
\newcommand*\VRL{\mathrm{V_{RL}}}
\newcommand*\MR{\mathrm{M_{R}}}
\newcommand*\tA{\mathrm{t_{0}}}
\newcommand*\xA{\mathrm{x_{0}}}
\let\oldcaption\caption
\long\def\caption#1{\oldcaption{\small\it #1}}
\begin{document}
\renewcommand{\figurename}{\it Fig.}
\graphicspath{{pictures/}}
\twocolumn
\maketitle
\begin{abstract}
Reasoned by its dynamical behavior, the memristor enables a lot of new applications in analog circuit design. Since some realizations are shown (e.g. 2007 by Hewlett Packard), the development of applications with memristors becomes more and more interesting. Whereas most of the research was done in the direction of memristor applications in neural networks and storage devices, less publications deal with practical applications of analog memristive circuits. But this topic is also promising further applications. Therefore, this article proposes a frequency dependent rectifier memristor bridge for different purposes (e.g. using as a programmable synaptic membrane voltage generator for Spike-Time-Dependent-Plasticity) and describes the circuit theory. In this context it is shown that the Picard Iteration is one possibility to solve the system of nonlinear state equations of memristor circuits analytically. An intuitive picture of how a memristor works in a network in general is given as well. In this context some research on the dynamical behavior of a HP memristor should be done. 
\end{abstract}
\section{Introduction}\label {Memsysteme}
\IEEEPARstart{T}{he} usage of memristors in analog circuit design enables new applications.
In \cite{pershin2011analog}, an ADC consisting of memristors has been proposed. Another applications are an automatic gain control circuit \cite{wey2010automatic} ,programmable analog circuits \cite{pershin2009practical}, an electrical potentiometer~\cite{pershin2009practical} or oscillators~\cite{bahgat2012memristor,talukdar2011memristor}. In \cite{merrikh2011memristor} memristors are used for basic arithmetic operations.

In this paper a frequency dependent rectifier memristor bridge is presented.
Therefore, a general description of memristive systems will be given first. Then the HP memristor is presented and some analysis about its dynamical behavior will be shown in~\ref{Time_response} and in~\ref{Properties} properties of memristors are given.
In~\ref{Graetz Schaltung mit Memristoren}, the memristor bridge is presented, which is solved analytically by Picard Iteration in\ref{PicardGraetz}.
Finally, the results of the present investigations are summarized. 

In General, a memristive system is described by the two equations \cite{di2009circuit}
\begin {equation}
V(t) = M(\underline{x},I,t)\cdot I(t) 
\end {equation}
\begin {equation}
\frac{\mathrm d\underline{x}}{\mathrm dt}=\underline{f}(\underline{x},I,t) 
\label{Zustandsgleichung}
\end {equation}
where $V(t)$ is the applied voltage and $I(t)$ is the current through the device. $M(\underline{x},I,t)$ is called memristance and relates voltage and current. Thus, similar to a linear resistor its dimension is Ohm. $\underline{x}$ is a vector of the dimension n which consists of the internal state variables.
The dynamics of this vector is specified by the state equation~\ref{Zustandsgleichung}.
\section{HP memristor}
The HP memristor was realized in 2007 by a team of researchers of Hewlett Packard around Stanley Williams~\cite{drakakis2010memristors, strukov2008missing}. While this memristor is passive, latest realization of a memristor is an active one on the base of niobium oxide~\cite{pickett2012sub}.
The HP memristor is made of a titanium dioxide layer which is located between two platinum electrodes. This layer is in the dimension of several nanometers and if there is an oxygen dis-bonding, its conductance will rise instantaneously. However, without doping, the layer behaves as an isolator. The area of oxygen dis-bonding will be referred to space-charge region and changes its extension if an electrical field is applied. This is done by drifting the charge carriers. The smaller the insulating layer the higher the conductance of the memristor. Also, the tunnel effect plays a crucial role. Without an external influence there is no change in the extension of the space-charge region. 
%
%
%
%
The internal state $x$ is the normed extent of the space-charge region and can be described by the equation
\begin {equation}
x:=\frac{w}{D}, \ 0\le x\le 1, \ x\in \mathbb{R} 
\end {equation}
where $\omega$ is the absolute extent of the space-charge region and $D$ is the absolute extent of the titanium dioxide layer.
%
%
The memristance can be described by the following equation \cite{strukov2008missing}:
\begin {equation}
M(x)=\Ron\cdot x+\Roff\cdot (1-x) \ .
\label{HPMemristanz}
\end {equation}
$\Ron$ is the resistance of the maximum conducting state and $\Roff$ represents the opposite case. The vector of internal state of the HP-memristor is one dimensional. That is why the scalar notation is used. 
The state equation is 
\begin {equation}
\frac{\mathrm dx}{\mathrm dt}=f(x,I,t)=\frac{\mu_v\cdot \Ron}{D^2}\cdot \IM(t) 
\label{Zustandsgleichung1}
\end {equation}
where $\mu_v$ is the oxygen vacancy mobility and $I_M(t)$ is the current through the device.
The following quantizations are used for memristors in this paper.
\begin{align*}
\Ron&=100\, \Omega, & \mu_v&=10^{-14}\, \frac{m^2}{s\cdot V}\\
\Roff&=16\, k\Omega, &  D&=10^{-8} \, m \ .
\end{align*}
These values are given as an example by the Hewlett Packard Team. The ratio of maximum and minimum value of the memristance can be as large as e.g. $160$. 
\subsection{Window function}
The extension of space-charge region is physically limited. Therefore, a window function shall be established to take account of saturation which is needed for a more realistic model. Several shapes of window functions are possible e.g.~\cite{biolek2009spice}. In the following, the window function 
\begin {equation}
\fW(x)=\left\{\begin{array}{cl} \left( 4\cdot x\cdot (1-x) \right.)^{\frac{1}{4}}, & \mbox{if}\ 0< x < 1 \\ 0, & \mbox{else} \end{array} \right.
\end {equation}
is used \cite{joglekar2009elusive}.
The current mathematical description of this window function shows a weakness in numerical calculations. 
Excluding external influences, in the case of $\{ x\le 0$ or  $x\ge 1\}$ the window function will be 0. Therefore, a state change of the system will not be possible. But considering $\VM$, the voltage over the memristor, for the two cases $x \le 0$ and $\VM>0$ or $x \ge 1$ and $\VM<0$, a state change should be possible. The window function will result in:
\begin {equation}
\fW(x,\VM)=\left\{\begin{array}{cl} (4\cdot x\cdot (1-x))^{\frac{1}{4}}, & \mbox{if}\ \epsilon< x < 1-\epsilon \\  & \mbox{or}\ x\le \epsilon \ and \ \VM>0 \\ & \mbox{or}\ x\ge 1-\epsilon \ and  \ \VM<0 \\ 0, & \mbox{else} \end{array} \right. \ .
\end {equation}
A better stability for numerical simulations is achieved by introducing $\epsilon$ and therefore an extended restriction of definition area. For all simulations in this paper $\epsilon$ equals $0.02$.
\subsection{Polarity of the HP memristor}\label{Symboldefinition}
%
An agreement concerning the memristor polarity will be introduced. The extension of the space-charge region is going together with a decrease of the memristance and vice versa. The polarity determines the behavior of the memristor. According to the direction of the applied field, if the memristance decreases the memristor will be ``forward biased". The other direction should be called ``reverse biased". Because of the minimum resistance and the time lag, both definitions are main intuitive. For a memristor the algebraic sign of $\VM$ equals that of $\IM$ and the polarity just depends on it.
Including the window function and the algebraic signs, Eq.~\ref{Zustandsgleichung1} extends to 
 \begin {equation}
      \frac{\mathrm dx}{\mathrm dt}=\frac{\mu_v\cdot\Ron}{D^2}  \cdot \fW\big(x,+\left(-\right)\VM\big)\cdot \big(+\left(-\right)\IM(t)\big) 
 \end {equation}
if the memristor is forward (reverse) biased.
\section{Time response of a HP memristor}\label{Time_response}
\subsection{Change of the internal state}
\begin{figure}[!t]
\centerline{\subfigure[current source: $I_0=0.0005A$, voltage source: $V_0=1V$.]{\includegraphics[width=1.6in]{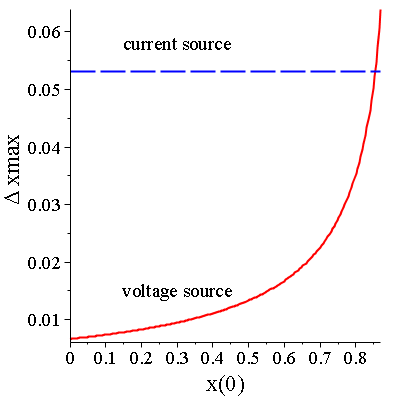}}
\hfil
\subfigure[current source: $I_0=0.0022A$, voltage source: $V_0=30 V$.]{\includegraphics[width=1.6in]{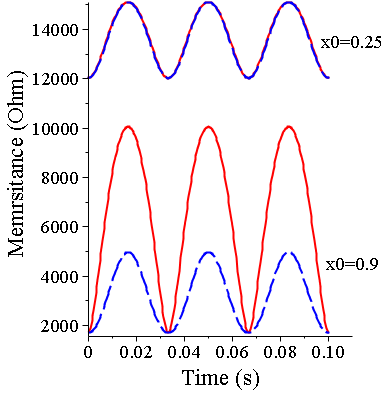}}}
\caption{{\textsl { Dashed line: supplied sine current source, solid line: supplied sine voltage source, $f=30 Hz$, (a) Maximal state change $\Delta x_{max}$ in dependency of $\xA$, $\Delta t=\frac{T}{2}$, (b) M(x) in dependency of t for two different initial states and two source types.}}}
\label{MemStrom}
\end{figure}

The calculations of this part will be done by neglecting the window function.
The aim is to investigate the time behavior of an AC sine current source $\IS$ which is connected in series with a HP memristor. 
Regarding the state equation~(\ref{Zustandsgleichung1})
\begin{equation}
\frac{\mathrm dx}{\mathrm dt}=\underbrace{\frac{\mu_v\cdot \Ron}{D^2}}_{=const}\cdot \underbrace{I_0\cdot \sin(\omega\cdot t)}_{\IS=\IM} \ ,
\end{equation}
dividing of the variables on both sides
\begin{equation}
\int_{\xA}^{\xA+\Delta x}{dx'}=const\cdot I_0\cdot \int_{\tA}^{\tA+\Delta t}{\sin(\omega\cdot t')dt'}, \ \Delta t\ge 0 
\end{equation}
and performing integration leads to
\begin{equation}
\Delta x=-\frac{I_0\cdot const}{\omega}\cdot \Big(\ cos\big(\omega\cdot (\tA+\Delta t)\big)-cos(\omega\cdot \tA)\Big) \ .
\label{Sinusstromquelle_Aenderungx}
\end{equation}
whereas $\xA:=x(\tA)$. At $(\tA+\Delta t)$ the extension of the space-charge region is $(\xA+\Delta x)$.
In the case of a memristor which is connected in series with a sine current source, the change of the space charge region $\Delta x$ does not depend on the initial state $\xA$ (see Eq.~\ref{Sinusstromquelle_Aenderungx}). 
There will be an other result if there is a sine voltage source instead.
In this case the state equation is
\begin{equation}
\frac{\mathrm dx}{\mathrm dt}=\underbrace{\frac{\mu_v\cdot\Ron}{D^2}}_{=const}\cdot \underbrace{\frac{V_0\cdot \sin (\omega\cdot t)}{M(x)}}_{\IM} \ .
\end{equation}
and because of $M(x)$ it leads by integration to a quadratic equation.
The relevant solution of this quadratic equation for $\Delta x$ with $\tA=0$ is
\begin{equation}
\Delta x=\left(r-\xA\right)-\sqrt{\left(r-\xA\right)^2+b\cdot \big(\cos(\omega\cdot\Delta t)-1\big)} 
\label {Deltax}
\end{equation}
whereas $r:=\frac{\Roff}{\Roff-\Ron}$ and $b:=\frac{2\cdot V_0\cdot const}{\omega\cdot (\Roff-\Ron)}$. 
The second solution is irrelevant because for $\Delta t=0$ the change of the space charge region $\Delta x$ has to be zero. Reasoned by the disregard of the window function, the usage of equation~(\ref{Deltax}) is bounded because the solution has to be within the physical limits. But the bottom line is that $\Delta x$ depends on the initial state $\xA$, if there is a voltage source in series with the memristor. As shown before, for a supplied current source it does not. Fig.~\ref{MemStrom}(a) illustrates these insight. Since the memristance is determined by the internal state, the change of the memristance behaves similarly. As you can see in figure~\ref{MemStrom}(b) for a sine voltage source the change of the memristance depends strongly on the internal memristance, however for a sine current source it does not. Note, in the case of a sinusoidal signal, the maximum change of the space charge region $\Delta x_{max}$ will be reached at the end of one half period. This is because the direction of the change of the space charge region depends on the algebraic sign of the external signal.
\subsection{Highest frequency for fully state change}\label{maximale Frequenz}
To estimate the dynamical behavior of the HP-memristor in circuits the frequency $\fcut$ for sinusoidal signals should be introduced. This is the highest frequency for which a memristor will be able to change from lowest to highest memristance or vice versa. It means that this state change completes exactly at the end of one half period ($\Delta\tcut = 0.5\cdot T_{cut}$). Note, for the used memristor model and same conditions, the required time for changing from a state A to a state B is the same as for the reverse process. $\fcut$ depends on the amplitude of the supplied source. Neglecting the window function the calculations should be performed for a single memristor in series with a supplied sine voltage source. This example allows rough estimates for circuits which are more complex.
Taking into account that $\omega$ equals $2\cdot \pi\cdot f$ and $T$ equals $\frac{1}{f}$. Integration of Eq.~\ref{Deltax} and solving for $f$ with $t=0$ leads to
\begin{equation}
\fcut=\frac{1}{\pi}\cdot\frac{V_0\cdot const}{\left(\frac {\Delta x^2}{2}+\xA\cdot \Delta x\right)\cdot\left(\Ron-\Roff\right)+\Delta x\cdot\Roff} 
\end{equation}
whereat $\fcut$ is direct proportional to the amplitude $V_0$.
\underline{example}:
Considering the restricted domain of $x$ the maximal frequency for switching from highest memristance $x(t=0)=0.02$ to lowest memristance $x(t=\frac{T}{2})=0.98$ should be calculated. Therefore $\Delta x$ is equal $0.96$. 
Calculating by using $V_0=30\,V$ leads to $ \fcut\approx 12.35675\, Hz$.
Testing by solving the state equation numerically leads to following results:\\
$x(t=\frac{1}{2\cdot f})=0.98$ for $f=12.35675\,Hz$\\
$x(t=\frac{1}{2\cdot f})=0.9424$ for $f=12.4\, Hz$\\
If the frequency is higher than $\fcut$, saturation wouldn't be reached. Considering the window function, the result of solving the state equation by simulation is $x(t=\frac{1}{2\cdot f})=0.98$ for $f=10.75\, Hz$. Therefore the calculations without window function are reasonable for rough estimates. Note, for $f> \fcut$ if the frequency is increasing the ratio of maximum and minimum value of the memristance will decrease. Note, for a supplied sine current source the calculation for $\fcut$ is also possible (Conversion of Eq.~\ref{Sinusstromquelle_Aenderungx}). As an example, for an amplitude $I_0=\frac{30 \, V}{\Roff}=1.9 \, mA$, $\fcut$ is about $6.2 \, Hz$.
\section{Properties}\label{Properties}
To summarize, the value of memristance depends on the current load of the past \cite{pershin2011memory}. If there is no current flow, the internal state will be retained. From this it follows that a memristor acts like a nonvolatile memory, whereas the range of values is continuous \cite{chua2011resistance,sinha2011evolving}. That is an interesting fact comparable to transistor memory technology. 
The difference between highest and lowest possible memristance is relative large \cite{strukov2008missing}. Indeed in the true sense, the memristor is no switch but it could be used for switching operations. Therefore, depending on the direction and the benchmark the memristor passes higher potentials and locks for lower ones.
Subject to the time shift the behavior of the memristor is frequency dependent. For $\lim_{f \rightarrow \infty}$ it behaves like a linear resistor \cite{di2009circuit} because the change of the internal state couldn't follow the rapid voltage change. For sufficiently small frequencies the nonlinearities are dominating whereas the time shift is direct proportional to the amplitude of the signal.
\section{Memristor bridge circuit}\label{Graetz Schaltung mit Memristoren}
\begin{figure}[!Hbtp]
\begin{minipage}[T]{0.27\textwidth}
\includegraphics[width=\textwidth]{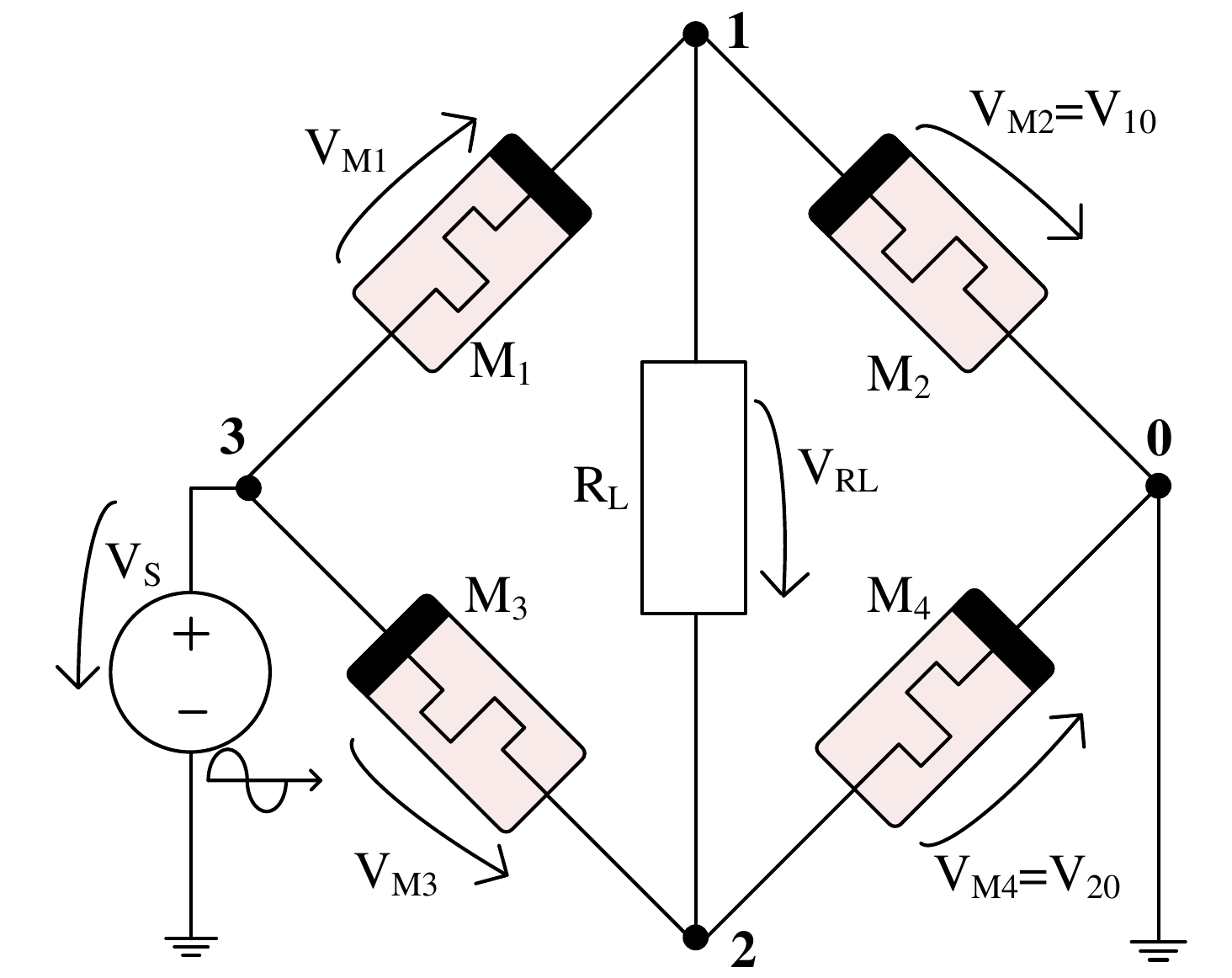}
\end{minipage}
\hfill
\begin{minipage}[T]{0.21\textwidth}
\begin{align}
\VM_1&=\VS-V_{10}\\
\VM_2&=V_{10}\\
\VM_3&=\VS-V_{20}\\
\VM_4&=V_{20}\\
\VRL&=V_{10}-V_{20}
\end{align} 
\vspace{0.01cm}
\end{minipage}
\caption{{\textsl {Schematic of the memristor bridge circuit.}}}
\label{Greatz}
\end{figure}

This circuit (Fig.~\ref{Greatz}) contains four HP memristors, one AC voltage source and one load resistor. Per definition for a positive voltage, memristors $M_{1,4}$ are forward biased, while $M_{2,3}$ are reverse biased.
In \cite{kimmemristor} and \cite{Cohen2012harmonics} this circuit is also been presented, but there are differences in application. While this paper deals with periodic signals and their specifics at different frequencies, in \cite{kimmemristor} pulses are used for synaptic weight programming. In \cite{Cohen2012harmonics} the focus lies on generation of nth-order harmonics and the effect of frequency doubling by using this circuit.
\subsection{Mathematical describtion}\label{Knotenspannungen}
%
At this point the new notation for the memristance 
\begin{equation}
M(x_n):=M_n, \ n\in \mathbb{N}
\end{equation}
should be established.
The voltages 
\begin{equation}
V_{10}:=V_{10}(x_1,x_2,x_3,x_4)=\VS\cdot \frac{\Num+M_2\cdot M_3 \cdot \RL}{\Den} 
\end{equation}
\begin{equation}
V_{20}:=V_{20}(x_1,x_2,x_3,x_4)=\VS\cdot \frac{\Num+M_1\cdot M_4 \cdot \RL}{\Den} 
\end{equation}
are defined from the nodes to ground, whereas the notations
\begin{align}
\Num \ :=& \ (M_1+M_3+\RL) \cdot M_2\cdot M_4 \\
\Den \ :=& \ \big(M_1+M_2\big)\cdot \big(M_3\cdot M_4+M_3\cdot \RL+M_4\cdot \RL\big) \notag \\ \ &+ M_1\cdot M_2 \cdot \big(M_3+M_4\big) \  .
\end{align}
are used for a better overview.
The structure of the circuit implies a nonlinear system of differential equations of the fourth order. The four state equations are 
\begin{align}
\frac{\mathrm dx_n}{\mathrm dt}=\frac{\mu_v\cdot\Ron}{D^2}\cdot \fW(x_n,\pm \VM_n)\cdot \underbrace{\pm \frac{\VM_n}{M(x_n)}}_{\IM_n}, \ & n \le 4 \ .
\end{align} 
For $x_{1,4}$ the sign is ``$+$" and for $x_{2,3}$ the sign is negative.
\subsection{Simulation results and functionality}
Using these state equations, numerical simulations are possible. From this point $\VS$ should be a sine AC voltage source and $R_L=1 \, k\Omega$.
%
%
\begin{figure}[!t]
\centerline{\subfigure[]{\includegraphics[width=1.6in]{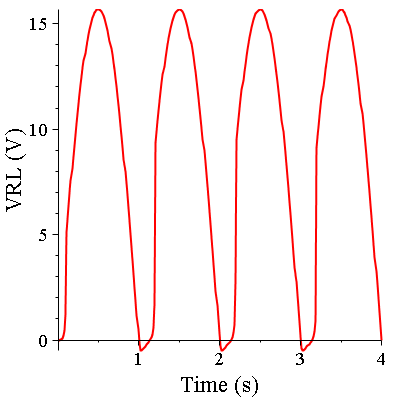}}
\hfil
\subfigure[]{\includegraphics[width=1.6in]{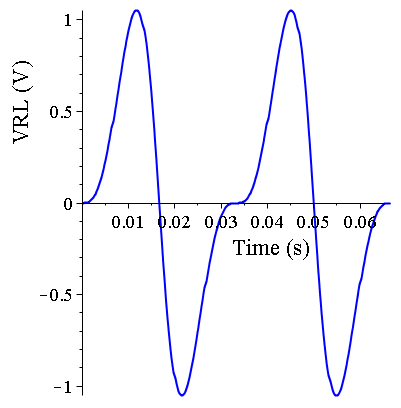}}}
\caption{{\textsl {Voltage $\VRL$ in dependency of time t, sine AC voltage source, $V_0=30\, V$, $x_n(t=0)=0.5$, (a) $f=0.5 \, Hz$,  (b) $f=30\, Hz$.}}}
\label{URL-Graetz-f-0,5}
\end{figure}

Regarding Fig.~\ref{URL-Graetz-f-0,5}, dependent on the frequency a qualitative difference for the voltage over the load $\VRL$ is detectable.
For low frequencies (represented by $f=0.5\, Hz$ in the case of $V_0=30\, V$) the output signal is almost exclusive positive, comparable with a rectifier circuit. Therefore, the frequency of the output signal is twice as high as the one of the input signal. Reasoned by the time shift of the change of state, which is subject to a HP memristor, there are temporary short negative peaks.
Using a higher excitation frequency (represented by $f=30\, Hz$ for $V_0=30\, V$) there is no rectifier function detectable. 

%
%
%
\begin{figure}[!t]
\centerline{\subfigure[]{\includegraphics[width=1.6in]{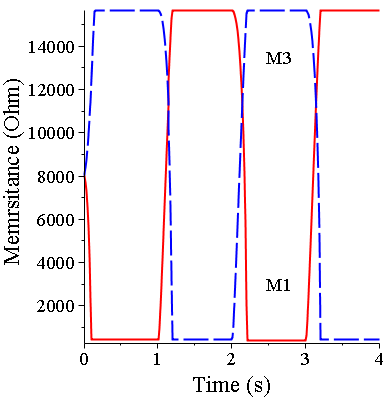}}
\hfil
\subfigure[]{\includegraphics[width=1.6in]{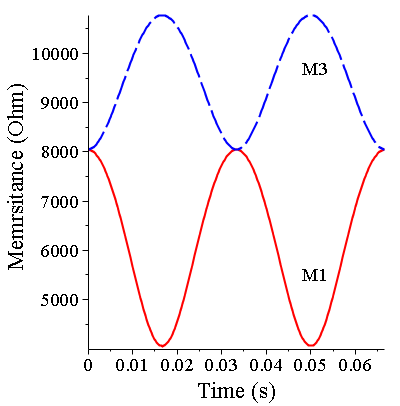}}}
\caption{{\textsl {Memristances $M_1$ and $M_3$ over time $t$, sine AC voltage source, $V_0=30\, V$, $x_n(t=0)=0.5$, (a) $f=0.5\, Hz$, (b) $f=30\, Hz$.}}}
\label{MemristanceM2-Graetz-f-0,5}
\end{figure}

By means of figure~\ref{MemristanceM2-Graetz-f-0,5}, this frequency selective behavior can be better understood.
For low frequencies ($f\le \fcut$), a complete state change will happen, the memristors are going into saturation. Beginning with an initial value $M_{1,4}$ is decreasing, while $M_{2,3}$ is increasing until saturation is reached. When the sine voltage source reaches its negative half period the process will be reversed until saturation is reached again and so on. During one period there is a change for the relational operator between both memristances. So every half period, the potential in node one is going to be higher than the potential in node two, which implies that $\VRL$ is always going to be positive.
For $f>\fcut$, the memristors do not completely switch from lowest to highest memristance. It follows that they are not going into saturation and the memristors will just reach the initial state after one period. So there is no change for the relational operator between $M_{1,4}$ and $M_{2,3}$ during one period and the relation of the memristances just depend on the initial conditions (See figure~\ref{MemristanceM2-Graetz-f-0,5}(b)).
For example, if the initial states are equal for all four memristors, then $M_{1,4}$ would be less than $M_{2,3}$. So $\VRL$ would be positive for the first half period too. But at the second half period, the Memristances $M_2$ and $M_3$ keep higher than $M_4$ and $M_1$. So the potential at node ``2" is higher than at node ``1", which implies that $\VRL$ is negative.
There are two reasons why the absolute value for the amplitude of $V_{RL}$ decreases for increasing frequency. On the one hand for equal initial conditions the maximum potential difference between node ``1" and ``2" decreases by increasing the frequency. On the other hand, dependent on the initial conditions, there is may a higher voltage drops across the memristors. Note, the maximum of the amplitude of $V_{RL}$ is not at $t=\frac{T}{4}$. At this time the amplitude of the sine source begins to decrease, but the change of the memristor states is continuing, which has firstly more impact.

The lower the frequencies the more the circuit behaves like a Graetz circuit. 
For very high frequencies the circuit is going to behave like a Wheatstone bridge. This can be proven by the voltage divider. For $\lim_{\RL \rightarrow \infty}$ it simply is
%
\begin{equation}
\VRL=V_{10}-V_{20}=\VS\cdot \left(\frac{M_2}{M_1+M_2}-\frac{M_4}{M_3+M_4}\right) \ ,
\end{equation}
and had to be zero for equal resistances, if the conclusion is true.
\begin{figure}[!t]
\centerline{\subfigure[$\lim_{f \rightarrow 0}$]{\includegraphics[width=1.6in]{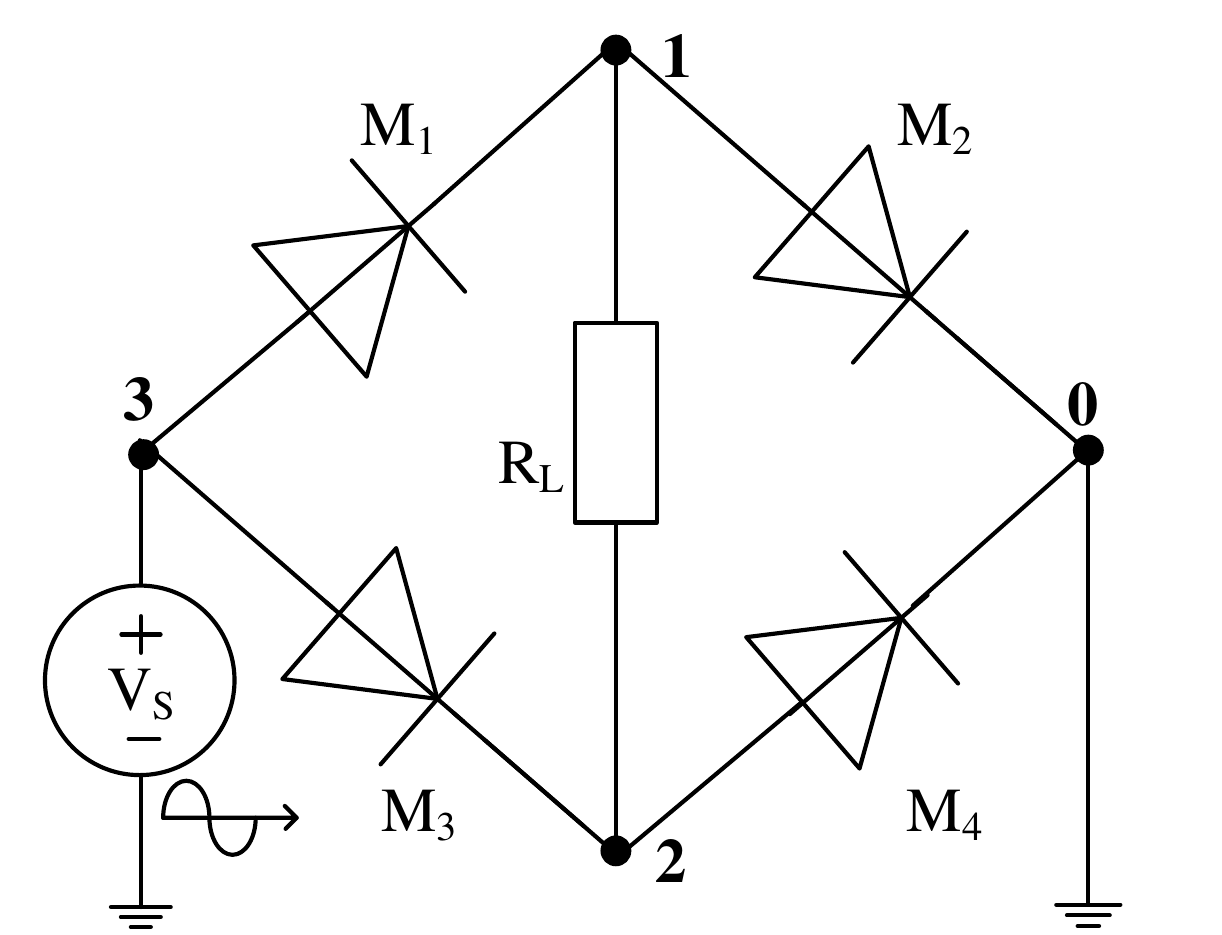}}
\hfil
\subfigure[$\lim_{f \rightarrow \infty}$]{\includegraphics[width=1.6in]{memristorbridge.pdf}}}
\caption{{\textsl {Approximate circuit equivalents: (a) Graetz circuit for very low frequencies, (b) Wheatstone circuit for very high frequencies.}}}
\end{figure}
Simulations illustrate that for e.g. $f=50\, kHz$ and $x_n(0)=0.5$ the amplitude of $V_{RL}$ equals $0.5 \, mV$, while the ratio between highest and lowest memristance of e.g. $M(x_1)$ equals $1.0002$.

\begin{figure}[!t]
\centerline{\subfigure[]{\includegraphics[width=1.6in]{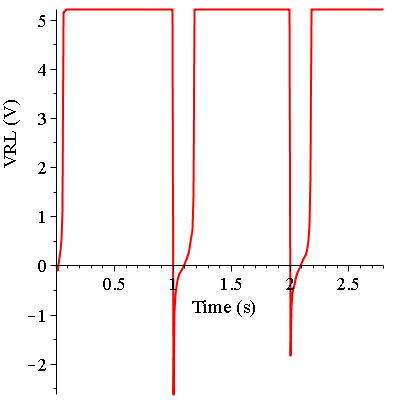}}
\hfil
\subfigure[]{\includegraphics[width=1.6in]{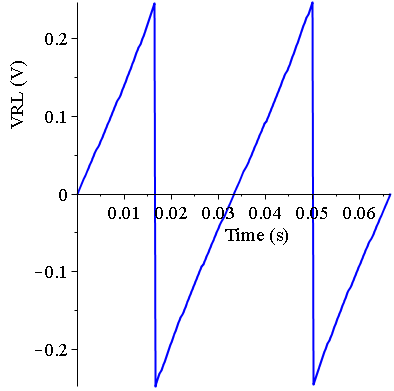}}}
\caption{{\textsl {$\VRL$ in dependency of t, supplied periodic square wave voltage source, $V_0=10\, V$, $x_n(t=0)=0.5$, (a) $f=0.5\,Hz$, (b) $f=30\,Hz$.}}}
\label{rechteck}
\end{figure}

\begin{figure}[!t]
\centerline{\subfigure[]{\includegraphics[width=1.6in]{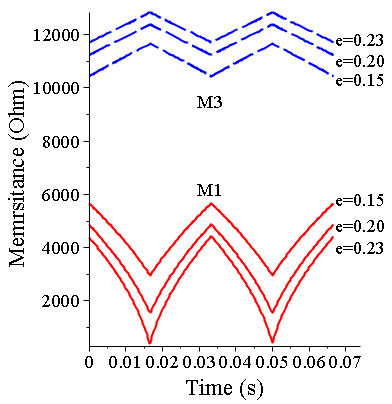}}
\hfil
\subfigure[]{\includegraphics[width=1.6in]{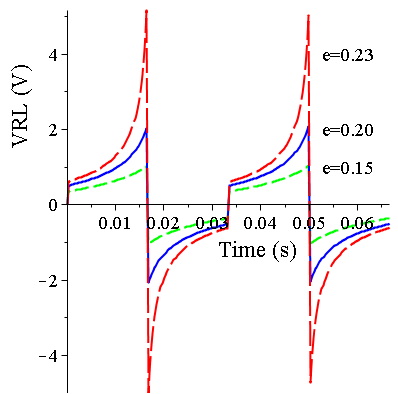}}}
\caption{{\textsl {(a) System supplied by a periodic square wave voltage source with $V_0=10\, V$, $f=30\,Hz$, $x_1(0)=x_4(0)=0.5+e$, $x_2(0)=x_3(0)=0.5-e$. (a) $M_1(x)$ and $M_3(x)$ in dependency of t for different initial conditions, (b) $\VRL$ in dependency of t for different initial conditions.}}}
\label{Spiketime}
\end{figure}

Using a supply periodic square wave voltage source with
\begin{equation}
\VS= sgn( \sin(\omega\cdot t)).
\end{equation} 
is another interesting example. The results for this are shown in Fig.~\ref{rechteck}. A frequency dependent behavior is also detectable. For higher frequencies the voltage over the load resistor is serrated. For low frequencies this voltage is almost constant with negative peaks.
The condition that the initial states of all memristors are equal leads to the in Fig.~\ref{rechteck} presented behavior. But what happens if the circuit is supplied by a periodic square wave voltage and the initial states are not equal? As it is described in chapter~\ref{Time_response} and shown in Fig.~\ref{Spiketime}(a), for a supplied periodic voltage and higher frequencies the variation of memristance depends on the initial conditions. So the voltage curve changes from the saw tooth form. As represented in Fig.~\ref{Spiketime}(b) it is possible to create a curve which is similar to synaptic membrane voltages which are used for Spike-Time-Dependent-Plasticity~\cite{linares2009memristance}. The amplitude of the voltage varies for different initial conditions, which could be set by a low frequency or DC voltage. 
%
%
\subsection{Possible applications}\label{Moegliche Anwendungen}
Depending on the frequency, there are two significant species. Fusing a common Graetz circuit and a Wheatstone circuit in one device is one application possibility. 
Replacing the independent source with a controlled one leads to a further application. The source is controlled by the load voltage and its frequency is variable. Keeping in mind, the amplitude of the output signal decreases by increasing the frequency. Because of that, a frequency controlled regulator is conceivable. If the amplitude exceeds a predefined value then the frequency should be increased to prevent a further amplification.
Using the circuit as a saw tooth generator is also possible.
Like mentioned before, the presented circuit also could be used as a programmable synaptic membrane voltage generator for Spike-Time-Dependent-Plasticity.
\section{Picard Iteration}\label{PicardGraetz}
At this point, the Picard Iteration should be introduced as a possibility to solve common memristive systems analytically.
One advantage related to e.g. the Volterra-series expansion is that the Picard Iteration converges more rapidly. In this chapter, the application possibility of this Iteration should be shown on the memristor bridge circuit. In favor, the general description of the Picard Iteration should be given first:
A nonlinear, dynamical system with
\begin{equation}
\frac{\mathrm dx}{\mathrm dt}=f\big( t,x(t)\big), \ x(t_0)=x_0
\end{equation}
is given, whereas $x(t)$ is Lipschitz continuous and an element of a Banach space.
Then, the iteration is given by
\begin{equation}
x^{[k+1]}:=x_0+\int_{t_0}^{t}{f\big(t^{'},x^{[k]}(t^{'})\big)\cdot \mathrm dt^{'}}, \ t\in \{t_0,t_0+\epsilon\} \ .
\end{equation}
whereas $k$ denotes the order of the iterative steps.

\begin{figure}[!hbtp]
\begin{minipage}[T]{0.27\textwidth}
\includegraphics[width=\textwidth]{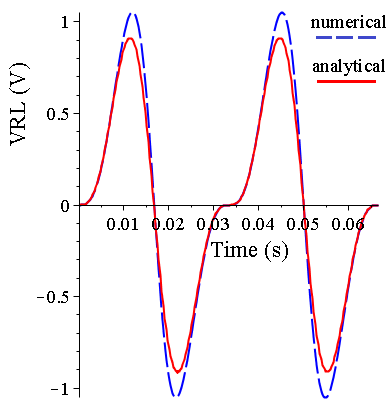}
\label{Graetzpicard}
\end{minipage}
\hfill
\begin{minipage}[T]{0.21\textwidth}
\small{\textit{Blue dashed line: numerical solution. \\ Red solid line: analytical solution by Picard Iteration (second iteration step). \\
 \\ Valid for both simulations: \\ $x_1^{[1]}=x_2^{[1]}=x_3^{[1]}=x_4^{[1]}=0.5$ and $V_0=30\, V$, $f=30\, Hz$. \\ \\}}
\end{minipage}
\caption{{\textsl {$\VRL$ over time t, analytical and numerical solution.}}}
\end{figure}

In the case of the memristor bridge circuit, $x(t)$ is the normed extent of the space-charge region.
The iteration should be performed without considering the window function. The window function suppresses changes of the state equation within the marginal area of the memristor. Therefore for $x_n^{[l+1]}$ the argument of the window function would be of the same iteration step. This would lead to a recursive mathematical expression and shows one weakness of the Picard Iteration. The iteration should be only performed for higher frequencies, thus, the window function can be neglected. By setting the initial conditions it should be noted that the internal states stay within the domain. Attention should be paid to the case $x>1$ because the memristance would become negative.
For memristive systems using the HP memristor model, the Picard iteration could be performed by
\begin{equation}
x_n^{[k+1]}:=x_0+\int_{t_0}^{t}{\frac{\mu_v\cdot \Ron}{D^2}\cdot \frac{\VM_n^{[k]}}{M(x_n^{[k]})} \cdot \mathrm dt^{'}}, \ t\in \{t_0,t_0+\epsilon\} 
\label{Picard2}
\end{equation}
with $\VM_n^{[k]}=\VM_n(t,x_1^{[k]},x_2^{[k]},...,x_n^{[k]})$.
Similar to chapter~\ref{Graetz Schaltung mit Memristoren} the used notation for the memristance $M(x_n^{[k]})$ is $M_n^{[k]}$ with $n,k\in \mathbb{N}$.
$\Num^{[k]}$ and $\Den^{[k]}$ should established as well. The only difference to their analogons in chapter~\ref{Graetz Schaltung mit Memristoren} is the usage of $M(x_n^{[k]})$ instead of $M(x_n)$.\\
%
%
\underline{1st iteration step}:
Using the initial states $x^{[1]}_n=x_n(0)$, $\frac{\mathrm dx^{[1]}_n}{\mathrm dt}=0$ as arguments for the first iteration step, then the equations for the currents are to be
\begin{align}
\IM_1^{[1]} \ & =\VS\cdot \underbrace{\frac{1}{M_1^{[1]}}\cdot \Bigg(1-\frac{\Num^{[1]}+M_2^{[1]}\cdot M_3^{[1]} \cdot \RL}{\Den^{[1]}}\Bigg)}_{\MR_1^{[1]}} \\
\IM_2^{[1]} \ & =\VS\cdot \underbrace{\frac{1}{M_2^{[1]}}\cdot \frac{\Num^{[1]}+M_2^{[1]}\cdot M_3^{[1]} \cdot \RL}{\Den^{[1]}}}_{\MR_2^{[1]}} \\
\IM_3^{[1]} \ & =\VS\cdot \underbrace{\frac{1}{M_3^{[1]}}\cdot \Bigg(1-\frac{\Num^{[1]}+M_1^{[1]}\cdot M_4^{[1]} \cdot \RL}{\Den^{[1]}}\Bigg)}_{\MR_3^{[1]}} \\
\IM_4^{[1]} \ & =\VS\cdot \underbrace{\frac{1}{M_4^{[1]}}\cdot \frac{\Num^{[1]}+M_1^{[1]}\cdot M_4^{[1]} \cdot \RL}{\Den^{[1]}}}_{\MR_4^{[1]}}  \ .
\end{align} 
\underline{2nd iteration step}:
\begin{equation}
\frac{\mathrm dx^{[2]}_n}{\mathrm dt}=\pm const\cdot V_0 \cdot \sin(\omega \cdot t) \cdot {\MR_n^{[1]}}
\end{equation} 
By dividing of the variables this equation is
\begin{equation}
\int_{x^{[1]}_n}^{x^{[2]}_n}{\mathrm dx^{'}}=\pm const\cdot V_0 \cdot \MR_n^{[1]}\cdot \int_{0}^{t}{\sin(\omega \cdot t^{'}) \mathrm dt^{'}} 
\end{equation} 
and leads by integration to
\begin{equation}
x^{[2]}_n=x^{[1]}_n\pm \frac{const\cdot V_0}{\omega} \cdot \MR_n^{[1]}\cdot (1-\cos(\omega \cdot t)) 
\end{equation} 
whereat for $x^{[2]}_{1,4}$ the sign is ``$+$" and for $x^{[2]}_{2,3}$ the sign is negative.
For this circuit for the second step the Picard Iteration gives a good quality solution (regarding Fig.~\ref{Graetzpicard}). Therefore no further iteration steps should be performed.
The voltage over the load resistor can be solved by
\begin{align}
\VRL^{[2]} 
\ & =\VS\cdot \frac{\RL}{\Den^{[2]}}\cdot (M_2^{[2]}\cdot M_3^{[2]}- M_1^{[2]}\cdot M_4^{[2]}) \ .
\end{align}
\section{conclusions}
At this paper two similar circuits consisting of HP memristors are presented. For both, a frequency selective behavior is presented. In contrast to higher frequencies for low frequencies there is a rectifier like behavior. This functional change with frequency is reasoned by the delay of changing of the internal state caused by an external source. To estimate the dynamical behavior in circuits, the time behavior of a single memristor in series with a periodic source was investigated. The configuration of the circuits leads to a nonlinear system of differential equations which describes the internal states. Using the Picard Iteration is one possibility to solve this system analytically. The frequency selective behavior can be used to realize two modes in one circuit (Graetz and Wheatstone circuit). Another applications are a frequency controlled regulator or a programmable synaptic membrane voltage generator for Spike-Time-Dependent-Plasticity.
\bibliographystyle{ieeetr}
\bibliography{Quellen}
\end{document}